\documentclass[11pt]{article}

\usepackage{geometry}
\usepackage{amsmath}
\usepackage{graphicx}
\usepackage{hyperref}
\usepackage{authblk}
\usepackage{setspace}
\usepackage{amssymb}
\usepackage{caption}
\usepackage{subcaption}

\usepackage{todonotes}

\usepackage[
backend=biber,
style=alphabetic, 
sorting=ynt
]{biblatex}
\title{Fast Geometric Embedding for \\ Node Influence Maximization}

\author[1]{Alexander Kolpakov}
\author[2]{Igor Rivin}  
\affil[1]{University of Austin, Austin TX, USA;
\texttt{akolpakov@uaustin.org}}
\affil[2]{Temple University, Philadelphia PA, USA;  
\texttt{rivin@temple.edu}}

\date{}  

\begin{document}

\maketitle

\begin{abstract}
Computing classical centrality measures such as betweenness and closeness is computationally expensive on large--scale graphs. In this work, we introduce an efficient force layout algorithm that embeds a graph into a low--dimensional space, where the radial distance from the origin serves as a proxy for various centrality measures. We evaluate our method on multiple graph families and demonstrate strong correlations with degree, PageRank, and paths--based centralities. As an application, it turns out that the proposed embedding allows one to find high--influence nodes in a network, and provides a fast and scalable alternative to the standard greedy algorithm.
\end{abstract}

\section{Introduction}
Graph centrality measures are a basic tool for quantifying the structural importance of vertices in a network \cite{freeman1979centrality, freeman1977set, brin1998, bonacich1972factoring}. They are used in settings ranging from information diffusion and social network analysis to biological interaction networks \cite{ws} and infrastructure modeling \cite{hk}. Classical centrality measures, however, are often expensive to compute on large graphs. In particular, path-based quantities such as betweenness \cite{freeman1977set, brandes2001} and closeness \cite{freeman1977set, freeman1979centrality} typically require repeated shortest-path computations, which become prohibitive as the size of the graph increases. This creates a practical need for scalable proxies that preserve the relative ordering of important vertices while avoiding the full combinatorial cost of exact centrality computation.

A natural approach is to seek low-dimensional graph representations from which node importance can be read off directly. Spectral methods provide informative global initializations, while force-directed methods are widely used to refine graph layouts by combining attraction and repulsion. In this paper, we combine these two ideas for a different purpose: not primarily visualization, but the construction of an embedding whose radial coordinate acts as a proxy for vertex importance.

More precisely, we propose a force layout algorithm that starts from a Laplacian embedding and then iteratively updates node positions by combining edge-based attraction with a repulsion term derived from intersecting edges. The resulting embedding is normalized after each iteration, and the radial distance from the origin is used as a scalar score. Empirically, this score exhibits a strong Spearman correlation with several standard centrality measures across multiple graph families and real-world datasets.

Our interest in this construction is twofold. First, it provides a computationally efficient surrogate for ranking nodes by importance. Second, it yields a practical heuristic for influence maximization: selecting nodes with large radial coordinate performs competitively with a greedy baseline in our experiments while requiring far less computation. Thus, the method is intended not as a replacement for exact centrality or exact influence optimization in every regime, but as a fast approximation procedure for large-scale settings in which classical combinatorial methods are too costly.

The main contribution of this paper is therefore a geometric embedding method tailored to node ranking. We formulate the algorithm, study some of its structural properties, and test it numerically on synthetic and real-world graphs. The theoretical discussion is partly heuristic, since a complete analysis appears to depend strongly on graph structure; nevertheless, the empirical behavior is consistent across a broad range of examples.

\section{Method Description}

\subsection{Spectral Initialization}
Let $G=(V,E)$ be a graph with $n=|V|$ vertices. We begin by computing a low-dimensional spectral embedding of $G$ from its graph Laplacian. Concretely, let $A$ be the adjacency matrix of $G$, let $D$ be the diagonal degree matrix, and let
\[
L = D-A
\]
be the combinatorial Laplacian. We choose an embedding dimension $d\geq 2$ and form the initial position matrix
\[
P^{(0)} = \begin{bmatrix} \phi_1 & \phi_2 & \cdots & \phi_d \end{bmatrix}^{\top} \in \mathbb{R}^{d\times n},
\]
where $\phi_1,\dots,\phi_d\in\mathbb{R}^n$ are Laplacian eigenvectors corresponding to the smallest nonzero eigenvalues of $L$.

Thus, for vertex $i\in V$, the initial position $p_i^{(0)}\in\mathbb{R}^d$ is obtained by taking the $i$-th coordinates of these eigenvectors. This gives a standard Laplacian layout, which serves as a globally informed initialization before the nonlinear refinement stage. In disconnected graphs, we apply the procedure to the largest connected component used in the benchmark.

\subsection{Force-Directed Refinement}
Starting from the spectral layout $P^{(0)}\in\mathbb R^{d \times n}$, we perform $T$ iterations of combined spring and intersection-repulsion updates. The attraction term keeps adjacent vertices at controlled distance, while the repulsion term discourages configurations in which many unrelated edges pass through the same region of the embedding.

\begin{enumerate}
  \item \emph{Spring Forces.}
  For each edge $e=(i,j)\in E$, let
  \[
    \Delta_{ij} = p_i - p_j,\qquad
    d_{ij} = \|\Delta_{ij}\| + \varepsilon,
  \]
  where $\varepsilon>0$ is a small constant. The elastic Hooke's force exerted on $i$ by $j$ is
  \[
    F^{\rm spring}_{ij}
      = -\,k_{\rm attr}\,\bigl(d_{ij}-L_{\min}\bigr)\,\frac{\Delta_{ij}}{d_{ij}}.
  \]
  Summing over all neighbors gives
  \[
    F^{\rm spring}_i = \sum_{j:(i,j)\in E} F^{\rm spring}_{ij}.
  \]

  \item \emph{Intersection Detection.}
  In dimension $d=2$, we test whether two edges $e=(u,v)$ and $f=(w,x)$ intersect using oriented-area predicates,
  \[
    \mathrm{orient}(a,b,c)
      = (b_x-a_x)(c_y-a_y) - (b_y-a_y)(c_x-a_x).
  \]
  The segments cross when
  \[
    \begin{aligned}
      &\mathrm{orient}(u,v,w)\cdot\mathrm{orient}(u,v,x)<0,\\
      &\text{and}\\
      &\mathrm{orient}(w,x,u)\cdot\mathrm{orient}(w,x,v)<0.
    \end{aligned}
  \]
  In dimensions $d>2$, the same term should be understood as penalizing near-crossings or geometric congestion after projection to a working two-dimensional plane (e.g. the first two coordinates of each vertex).

  \item \emph{Repulsion at Crossings.}
  For each intersecting pair $(e,f)$, we define the average midpoint
  \[
    m_{e,f} = \frac{p_u+p_v+p_w+p_x}{4}.
  \]
  We then apply to each endpoint $i\in\{u,v,w,x\}$ the repulsive contribution
  \[
    F^{\rm intersect}_{i,e,f}
      = k_{\rm inter}\,\frac{p_i-m_{e,f}}{\|p_i-m_{e,f}\|^2+\varepsilon}.
  \]

  Summing over all crossing pairs yields
  \[
    F^{\rm intersect}_i \;=\;\sum_{(e,f)\,\text{crossing}} F^{\rm intersect}_{i,e,f}.
  \]

  \item \emph{Position Update and Normalization.}  
  We combine the two force components and then re-‐center and rescale:
  \[
    P = P^{(t)} + F^{\rm spring} + F^{\rm intersect},
  \]
  \[
    P^{(t+1)}
      = \frac{\,P - \mu\,}
             {\sigma + \varepsilon},
  \]
  where
  \[
    \mu \;=\;\frac1n\sum_{i=1}^n p_i,\quad
    \sigma \;=\;\sqrt{\frac1n \sum_{i=1}^n \|p_i - \mu\|^2},
  \]
  which prevents uncontrolled drift and stretching.
\end{enumerate}

\section{Theoretical Analysis}

In this section we discuss some structural properties of the proposed algorithm and give heuristic reasons why the radial coordinate may correlate with node importance for broad classes of graphs. It seems that a detailed analysis is only possible under some assumptions that impose the graph structure, while we would rather opt for a more general and possibly more heuristic argument. Nevertheless, it seems that the algorithm shows consistently strong results on both synthetic and real world datasets of various kinds, thus showing its practical tenability. 

\subsection{Fixed-point existence}

Let $P=(p_1,\dots,p_n)\in \mathbb{R}^{d\times n}$ denote a configuration of node positions. Define the force-update map
\[
\mathcal{F}(P)=P+F^{\rm spring}(P)+F^{\rm intersect}(P),
\]
and the normalization map
\[
\mathcal{N}(P)=\frac{P-\mu(P)}{\sigma(P)+\varepsilon},
\]
where
\[
\mu(P)=\frac1n\sum_{i=1}^n p_i,
\qquad
\sigma(P)=\sqrt{\frac1n\sum_{i=1}^n \|p_i-\mu(P)\|^2},
\]
and $\varepsilon>0$ is fixed. The full iteration is therefore
\[
P^{(t+1)}=\mathcal{T}(P^{(t)}),\qquad \mathcal{T}:=\mathcal{N}\circ \mathcal{F}.
\]

We now explain why $\mathcal{T}$ has a fixed point on a natural compact configuration space. Consider the set
\[
K=\left\{P\in\mathbb{R}^{d\times n}:
\mu(P)=0,\;
\frac1n\sum_{i=1}^n \|p_i\|^2\leq 1
\right\}.
\]
The set $K$ is convex, closed, and bounded, hence compact in any finite dimension $n$.

By construction, for every configuration $P$ for which $\mathcal{F}(P)$ is defined, the normalized configuration $\mathcal{T}(P)$ satisfies
\[
\mu(\mathcal{T}(P))=0.
\]
Moreover, if we write $Q=\mathcal{F}(P)$, then
\[
\mathcal{T}(P)=\frac{Q-\mu(Q)}{\sigma(Q)+\varepsilon},
\]
so
\[
\frac1n\sum_{i=1}^n \|\mathcal{T}(P)_i\|^2
=
\frac{\sigma(Q)^2}{(\sigma(Q)+\varepsilon)^2}
\leq 1.
\]
Hence $\mathcal{T}(K)\subseteq K$.

It remains to note that, away from degenerate configurations where denominators vanish, both $F^{\rm spring}$ and $F^{\rm intersect}$ depend continuously on $P$ because of the regularization by $\varepsilon$. Therefore $\mathcal{T}$ is continuous on $K$. Brouwer's fixed-point theorem now implies that $\mathcal{T}$ has at least one fixed point in $K$.

Thus, there exists a normalized configuration $P^\ast\in K$ such that
\[
\mathcal{T}(P^\ast)=P^\ast.
\]
This does not by itself imply uniqueness or convergence from every initialization, but it does show that the iteration admits at least one equilibrium configuration in the normalized state space.

\subsection{Energy and constrained equilibrium}

The force terms used in the algorithm suggest the following energy functional:
\[
\begin{aligned}
        \mathcal{E}(P) = \frac{k_{\text{attr}}}{2} \sum_{(i,j)\in E} \left( \|p_i - p_j\| - L_{\min} \right)^2 \,\,+ \sum_{(e,f)\,\in \mathcal{I}(i)}  \frac{k_{\text{inter}}}{\|p_i - m_{e,f}\|^2},
\end{aligned}
\]
where 
\[
\begin{aligned}
    \mathcal{I}(i) = \{ (e,f) \in E\times E \;|\; e = (i,j),\; &f = (u,v),\\ &\text{$e$ and $f$ intersect.}  \}
\end{aligned}
\]
and
\[
m_{e,f}=\frac{p_u+p_v+p_w+p_x}{4}.
\]

The first term penalizes deviations of edge lengths from the preferred scale $L_{\min}$, while the second penalizes geometric congestion near edge crossings. The normalization step of the algorithm imposes the constraints
\[
\sum_{i=1}^n p_i=0,
\qquad
\frac1n\sum_{i=1}^n \|p_i\|^2=1.
\]
Accordingly, one may consider the constrained variational problem for
\[
\mathcal{L}(P,\lambda,\alpha)
=
\mathcal{E}(P)
-
\lambda\cdot \sum_{i=1}^n p_i
-
\alpha\left(\frac1n\sum_{i=1}^n \|p_i\|^2-1\right),
\]
where $\lambda\in\mathbb{R}^d$ and $\alpha\in\mathbb{R}$ are Lagrange multipliers.

Stationarity with respect to $p_i$ gives
\[
\nabla_{p_i}\mathcal{E}(P)-\lambda-\frac{2\alpha}{n}p_i=0.
\]
Summing over $i$ and using translation invariance of $\mathcal{E}$ yields
\[
\sum_{i=1}^n \nabla_{p_i}\mathcal{E}(P)=0,
\]
hence
\[
0
=
n\lambda+\frac{2\alpha}{n}\sum_{i=1}^n p_i
=
n\lambda,
\]
because the centering constraint gives $\sum_i p_i=0$. Therefore
\[
\lambda=0,
\]
and the stationarity condition reduces to
\[
\nabla_{p_i}\mathcal{E}(P)=\frac{2\alpha}{n}p_i.
\]

This identity has a simple geometric interpretation: at a constrained critical point, the net energy gradient at node $i$ is parallel to the radial direction $p_i$. In other words, the tangential component of the gradient vanishes, and only the radial component survives because of the variance constraint. Thus the normalization step biases the dynamics toward configurations in which positional adjustments are reflected primarily in the radial coordinate.

This observation does not by itself yield explicit formulas for the positions $p_i$. Rather, it indicates that the radial direction plays a distinguished role, and allows us to write the following heuristic equalities (at least, after sufficiently many steps): 

\[
    p_i = p_j + c_{ij}\, d_{ij}\, \mathbf{r}_i + k_{ij} \mathbf{r}_i^\perp,
\]
\[
    p_i = m_{e,f} + c_{e,f}\, \|p_i\| \mathbf{r}_i + l_{ij} \mathbf{r}_i^\perp,
\]
for some $c_{ij} > 0$ and $c_{e,f} > 0$, $\mathbf{r}_i = \frac{p_i}{\|p_i\|}$ the unit vector pointing at $p_i$ (i.e., the radial direction towards $p_i$), and $d_{ij} = \|p_i - p_j\|$ the distance between two nodes. 

\subsection{Heuristic scaling relation between radial distance and degree}

Now let us turn to a heuristic argument suggesting why $\|p_i\|$ may correlate with degree, and more generally with centrality-like quantities.

At equilibrium, the force balance for node $i$ is
\[
F^{\rm spring}_i+F^{\rm intersect}_i=0.
\]
Writing this out gives
\[
k_{\rm attr}\sum_{j:(i,j)\in E}
\bigl(d_{ij}-L_{\min}\bigr)\frac{p_i-p_j}{d_{ij}}
=
k_{\rm inter}
\sum_{(e,f)\in \mathcal{I}(i)}
\frac{p_i-m_{e,f}}{\|p_i-m_{e,f}\|^2+\varepsilon},
\]
where $\mathcal{I}(i)$ denotes the set of intersecting edge pairs whose repulsion term acts on node $i$ (already defined above).

To obtain a tractable approximation, suppose that near equilibrium the dominant components of both sides are approximately radial. This is consistent with the constrained optimization discussion above. 

Putting $\delta_{i} = \frac{1}{{\rm deg}(i)} \sum_j c_{ij}\, (d_{ij} - L_{min})$, we get 
\[
    k_{\rm attr}\,  {\rm deg}(i)\; \delta_i \;\mathbf{r}_i = \frac{k_{\rm inter}}{\|p_i\|}\, \sum_{(e,f) \in \mathcal{I}(i)} \frac{1}{ c_{e,f}} \; \mathbf{r}_i,
\]
which is equivalent to the scalar equality 
\[
    \|p_i\| = \frac{k_{\rm inter}}{k_{\rm attr}\, \mathrm{deg}(i)\, \delta_i} \, \sum_{(e,f) \in \mathcal{I}(i)} \frac{1}{ c_{e,f}}
\]

This suggests introducing the intersection-load quantity
\[
\mathrm{int}(i):=
\sum_{(e,f)\in\mathcal{I}(i)} \frac{1}{c_{e,f}},
\]
so that
\[
\|p_i\|
\approx
\frac{k_{\rm inter}}{k_{\rm attr}\,\deg(i)\,\delta_i}\,
\mathrm{int}(i).
\]
Thus the radial coordinate is predicted to be large when node $i$ is incident to many geometrically congested interactions relative to its local elastic scale.

To connect this with degree, one may posit a graph-dependent scaling law
\[
\mathrm{int}(i)\approx C\,\deg(i)^{1+\beta},
\]
with constants $C>0$ and $\beta\geq 0$ varying by graph family. If, in addition, $\delta_i$ does not vary too strongly across nodes, then
\[
\|p_i\|\approx C'\deg(i)^\beta
\]
for another constant $C'>0$. This would imply a positive rank correlation between $\|p_i\|$ and degree.

This, however, is not a theorem, but rather a heuristic argument. For instance, if edges in the neighborhood of $i$ have approximately comparable intersection probability $\rho$, and neighboring degrees are of the same order as $\deg(i)$, then one expects the number of nearby crossing interactions to grow superlinearly in $\deg(i)$, often closer to quadratic than linear. In that regime, $\mathrm{int}(i)$ naturally increases with degree, which in turn pushes $\|p_i\|$ outward. Thus, our assumptions are largely plausible. 

This also helps explain why the method may correlate with other centrality measures. In many graph families, such centrality measures as degree, PageRank, and betweenness,  are positively correlated \cite{valente2008correlated}. On the other hand, the approximation may fail for structurally exceptional vertices, such as low-degree bridge vertices with unusually high betweenness. This implies that in some graph classes more delicate structural roles of certain edges may be less faithfully represented by our embedding.

\section{Achieving efficiency}
In order to achieve computational efficiency, we reduce the computation of intersection forces $F^{\rm intersect}$ to an analogous computation on the $k$ nearest neighbor ($kNN$) graph of midpoints of intersecting edges. This avoids overwhelming computational load if the number of intersections is much larger than $O(E)$. 

The $kNN$ graph can be efficiently computed, especially if $k$ is not very large, and a theoretical algorithm would take about $O(E^{1+\tau})$, steps, with $\tau \in (0,1)$ \cite{chen2009fast}. Also, approximate $kNN$ methods are known to be very efficient in practice \cite{nndescent}. However, subsampling the midpoints together with GPU/TPU acceleration \cite{jax} makes the runtime much shorter. After that, the repulsion forces $F^{\rm intersect}$ are computed in $O(kE)$ steps. 

The attraction forces $F^{\rm spring}$ are naturally provided by the graph connectivity structure, and take $O(E)$ steps to compute. 

\section{Numerical Experiments}
We evaluated our method on several synthetic graph families, including Erd\"os-Renyi \cite{er}, Watts--Strogatz \cite{ws}, power law cluster \cite{hk}, and balanced tree graphs, as well as real world datasets. For each graph, we compute standard centrality measures: degree, betweenness \cite{freeman1979centrality, brandes2001}, eigenvector \cite{bonacich1972factoring}, PageRank \cite{brin1998}, closeness \cite{freeman1977set, freeman1979centrality}, and node load \cite{goh2001universal}, and compare them with the radial distances from our embedding using Spearman correlation. We also provide its confidence interval with confidence level $\gamma = 0.95$, obtained by bootstrapping.

\subsection{Experimental Setup}
The force layout is computed using a JAX--based implementation \cite{graphem-github} with a specified number of iterations, attraction and repulsion parameters, and a user--defined embedding dimension. Correlation coefficients are calculated between the radial distance and each centrality measure. In practice, our algorithm shows strong correlations in hub--dominated networks, while preserving rank order in hierarchical structures.

\subsection{Synthetic Datasets}
Tables~\ref{tab:er2}--\ref{tab:er4} summarize the correlations for Erd\"os--Renyi graphs \cite{er}, Tables~\ref{tab:ws2}--\ref{tab:ws4} are  for Watts--Strogatz graphs \cite{ws}, and Tables~\ref{tab:powerlaw2}--\ref{tab:powerlaw4} for power law cluster graphs \cite{hk}. These results demonstrate that in small--world and scale--free networks, the radial distance strongly approximates global centrality in the sense of the ordering of nodes by centrality. 

Here we use Spearman's $\rho$ correlation instead of Pearson's correlation as the relationship between radial ordering and centrality is not necessarily linear (as the force layout is highly non--linear), and because what matters most is the ordering, not the actual distance or centrality values. 

In all cases we can observe that embeddings in dimension $2$ already demonstrate good correlation between the radial distance and centrality measures. However, as the embedding dimension growth up to $4$, all correlations become stronger. There could be a slight increase in correlations after dimension $4$: however, it is on the order of magnitude smaller than between dimension $2$ and $4$. More experimental data is available on GitHub \cite{graphem-github}. 

On the other hand, being a planar graph the balanced ternary tree shows best correlations between the radial distance in the embedding and vertex centralities in dimension $2$, see Table~\ref{tab:tree2}. These correlations visibly deteriorate in higher dimensions, as shown in Table~\ref{tab:tree4}. 

\begin{table}[h!]
\centering
\begin{tabular}{lccc}
\hline
\textbf{Centrality Measure} & \boldmath$\rho$ & \textbf{95\% CI} & \boldmath$p$ \\
\hline
Degree       & 0.829 & [0.803, 0.854] & $< 10^{-6}$ \\
Betweenness  & 0.845 & [0.817, 0.867] & $< 10^{-6}$ \\
Eigenvector  & 0.806 & [0.778, 0.833] & $< 10^{-6}$ \\
PageRank     & 0.835 & [0.807, 0.859] & $< 10^{-6}$ \\
Closeness    & 0.830 & [0.802, 0.855] & $< 10^{-6}$ \\
Node Load    & 0.845 & [0.818, 0.866] & $< 10^{-6}$ \\
\hline
\end{tabular}
\caption{Spearman correlations of centrality measures with the radial distance in graph embeddings for Erd\"os -- Renyi graphs. Embedding dimension $2$.}
\label{tab:er2}
\end{table}

\begin{table}[h!]
\centering
\begin{tabular}{lccc}
\hline
\textbf{Centrality Measure} & \boldmath$\rho$ & \textbf{95\% CI} & \boldmath$p$ \\
\hline
Degree       & 0.963 & [0.956, 0.968] & $< 10^{-6}$ \\
Betweenness  & 0.966 & [0.959, 0.971] & $< 10^{-6}$ \\
Eigenvector  & 0.948 & [0.939, 0.955] & $< 10^{-6}$ \\
PageRank     & 0.965 & [0.959, 0.970] & $< 10^{-6}$ \\
Closeness    & 0.963 & [0.956, 0.968] & $< 10^{-6}$ \\
Node Load    & 0.966 & [0.960, 0.970] & $< 10^{-6}$ \\
\hline
\end{tabular}
\caption{Spearman correlations of centrality measures with the radial distance in graph embeddings for Erd\"os -- Renyi graphs. Embedding dimension $4$.}
\label{tab:er4}
\end{table}

\begin{table}[h!]
\centering
\begin{tabular}{lccc}
\hline
\textbf{Centrality Measure} & \boldmath$\rho$ & \textbf{95\% CI} & \boldmath$p$ \\
\hline
Degree       & 0.896 & [0.877, 0.912] & $< 10^{-6}$ \\
Betweenness  & 0.748 & [0.718, 0.776] & $< 10^{-6}$ \\
Eigenvector  & 0.646 & [0.605, 0.682] & $< 10^{-6}$ \\
PageRank     & 0.897 & [0.878, 0.912] & $< 10^{-6}$ \\
Closeness    & 0.594 & [0.549, 0.633] & $< 10^{-6}$ \\
Node Load    & 0.743 & [0.711, 0.771] & $< 10^{-6}$ \\
\hline
\end{tabular}
\caption{Spearman correlations of centrality measures with the radial distance in graph embeddings for Watts--Strogatz graphs. Embedding dimension $2$.}
\label{tab:ws2}
\end{table}

\begin{table}[h!]
\centering
\begin{tabular}{lccc}
\hline
\textbf{Centrality Measure} & \boldmath$\rho$ & \textbf{95\% CI} & \boldmath$p$ \\
\hline
Degree       & 0.956 & [0.948, 0.962] & $< 10^{-6}$ \\
Betweenness  & 0.810 & [0.784, 0.831] & $< 10^{-6}$ \\
Eigenvector  & 0.711 & [0.676, 0.742] & $< 10^{-6}$ \\
PageRank     & 0.947 & [0.937, 0.955] & $< 10^{-6}$ \\
Closeness    & 0.656 & [0.621, 0.691] & $< 10^{-6}$ \\
Node Load    & 0.804 & [0.779, 0.826] & $< 10^{-6}$ \\
\hline
\end{tabular}
\caption{Spearman correlations of centrality measures with the radial distance in graph embeddings for Watts--Strogatz graphs. Embedding dimension $4$.}
\label{tab:ws4}
\end{table}

\begin{table}[h!]
\centering
\begin{tabular}{lccc}
\hline
\textbf{Centrality Measure} & \boldmath$\rho$ & \textbf{95\% CI} & \boldmath$p$ \\
\hline
Degree       & 0.783 & [0.750, 0.814] & $< 10^{-6}$ \\
Betweenness  & 0.635 & [0.590, 0.678] & $< 10^{-6}$ \\
Eigenvector  & 0.411 & [0.353, 0.463] & $< 10^{-6}$ \\
PageRank     & 0.805 & [0.773, 0.832] & $< 10^{-6}$ \\
Closeness    & 0.352 & [0.297, 0.407] & $< 10^{-6}$ \\
Node Load    & 0.644 & [0.598, 0.692] & $< 10^{-6}$ \\
\hline
\end{tabular}
\caption{Spearman correlations of centrality measures with the radial distance in graph embeddings for power law cluster graphs. Embedding dimension $2$.}
\label{tab:powerlaw2}
\end{table}

\begin{table}[h!]
\centering
\begin{tabular}{lccc}
\hline
\textbf{Centrality Measure} & \boldmath$\rho$ & \textbf{95\% CI} & \boldmath$p$ \\
\hline
Degree       & 0.866 & [0.841, 0.888] & $< 10^{-6}$ \\
Betweenness  & 0.733 & [0.696, 0.766] & $< 10^{-6}$ \\
Eigenvector  & 0.561 & [0.516, 0.600] & $< 10^{-6}$ \\
PageRank     & 0.869 & [0.845, 0.892] & $< 10^{-6}$ \\
Closeness    & 0.500 & [0.451, 0.546] & $< 10^{-6}$ \\
Node Load    & 0.738 & [0.700, 0.770] & $< 10^{-6}$ \\
\hline
\end{tabular}
\caption{Spearman correlations of centrality measures with the radial distance in graph embeddings for power law cluster graphs. Embedding dimension $4$.}
\label{tab:powerlaw4}
\end{table}

\begin{table}[h!]
\centering
\begin{tabular}{lccc}
\hline
\textbf{Centrality Measure} & \boldmath$\rho$ & \textbf{95\% CI} & \boldmath$p$ \\
\hline
Degree       & 0.816 & [0.810, 0.822] & $< 10^{-6}$ \\
Betweenness  & 0.772 & [0.769, 0.774] & $< 10^{-6}$ \\
Eigenvector  & 0.670 & [0.660, 0.680] & $< 10^{-6}$ \\
PageRank     & 0.830 & [0.823, 0.837] & $< 10^{-6}$ \\
Closeness    & 0.772 & [0.770, 0.775] & $< 10^{-6}$ \\
Node Load    & 0.772 & [0.769, 0.775] & $< 10^{-6}$ \\
\hline
\end{tabular}
\caption{Spearman correlations of centrality measures with the radial distance in graph embeddings for balanced ternary trees. Embedding dimension $2$.}
\label{tab:tree2}
\end{table}

\begin{table}[h!]
\centering
\begin{tabular}{lccc}
\hline
\textbf{Centrality Measure} & \boldmath$\rho$ & \textbf{95\% CI} & \boldmath$p$ \\
\hline
Degree       & 0.479 & [0.462, 0.497] & $< 10^{-6}$ \\
Betweenness  & 0.463 & [0.446, 0.480] & $< 10^{-6}$ \\
Eigenvector  & 0.428 & [0.412, 0.446] & $< 10^{-6}$ \\
PageRank     & 0.477 & [0.460, 0.495] & $< 10^{-6}$ \\
Closeness    & 0.463 & [0.446, 0.480] & $< 10^{-6}$ \\
Node Load    & 0.463 & [0.446, 0.480] & $< 10^{-6}$ \\
\hline
\end{tabular}
\caption{Spearman correlations of centrality measures with the radial distance in graph embeddings for balanced ternary trees. Embedding dimension $4$.}
\label{tab:tree4}
\end{table}

\begin{figure}[htbp]
    \centering
    \subfloat[Initial Laplacian Embedding]{%
        \includegraphics[width=0.45\linewidth]{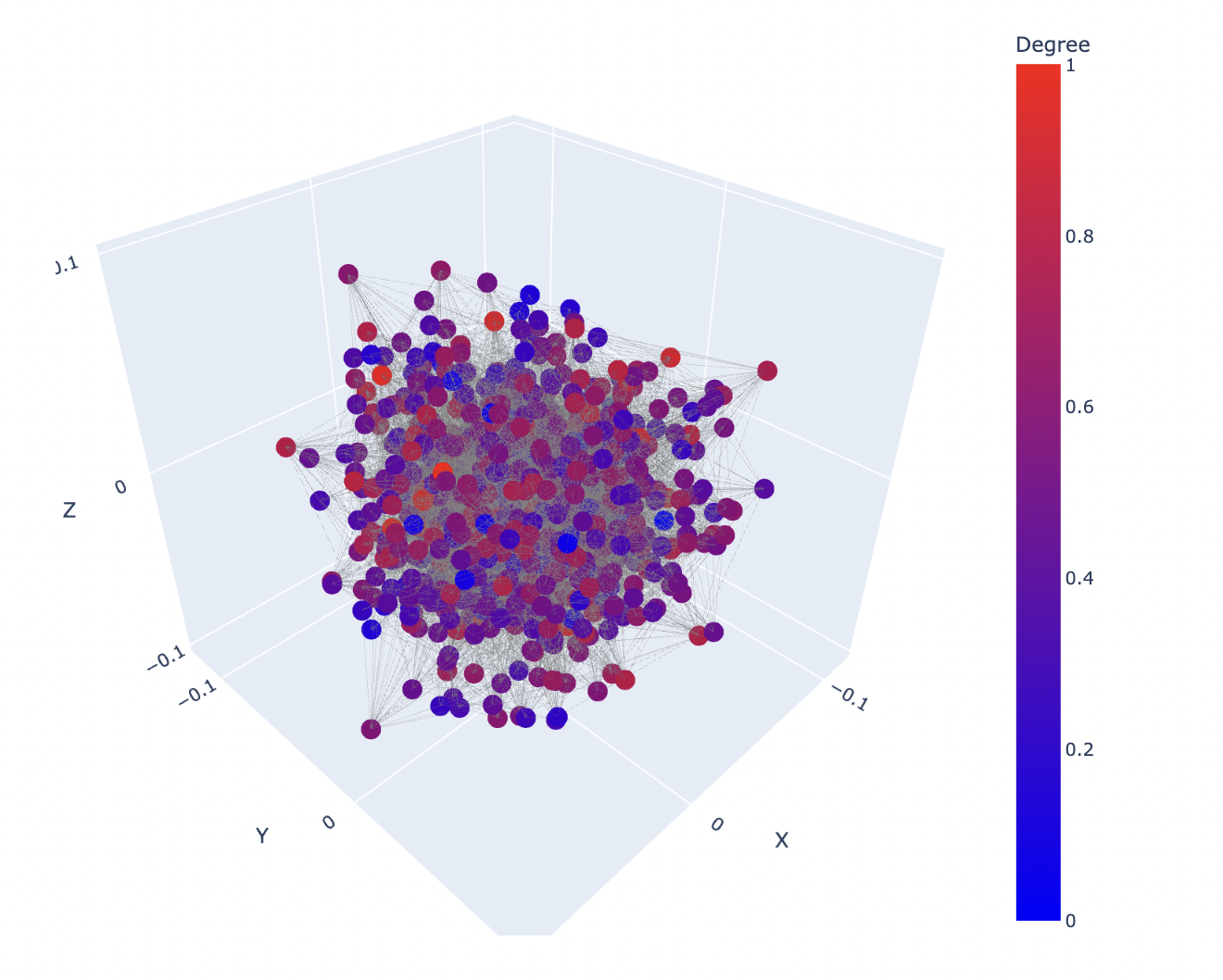}%
        \label{fig:initial_embed}
    }
    \hfill
    \subfloat[Finalized Force Layout]{%
        \includegraphics[width=0.45\linewidth]{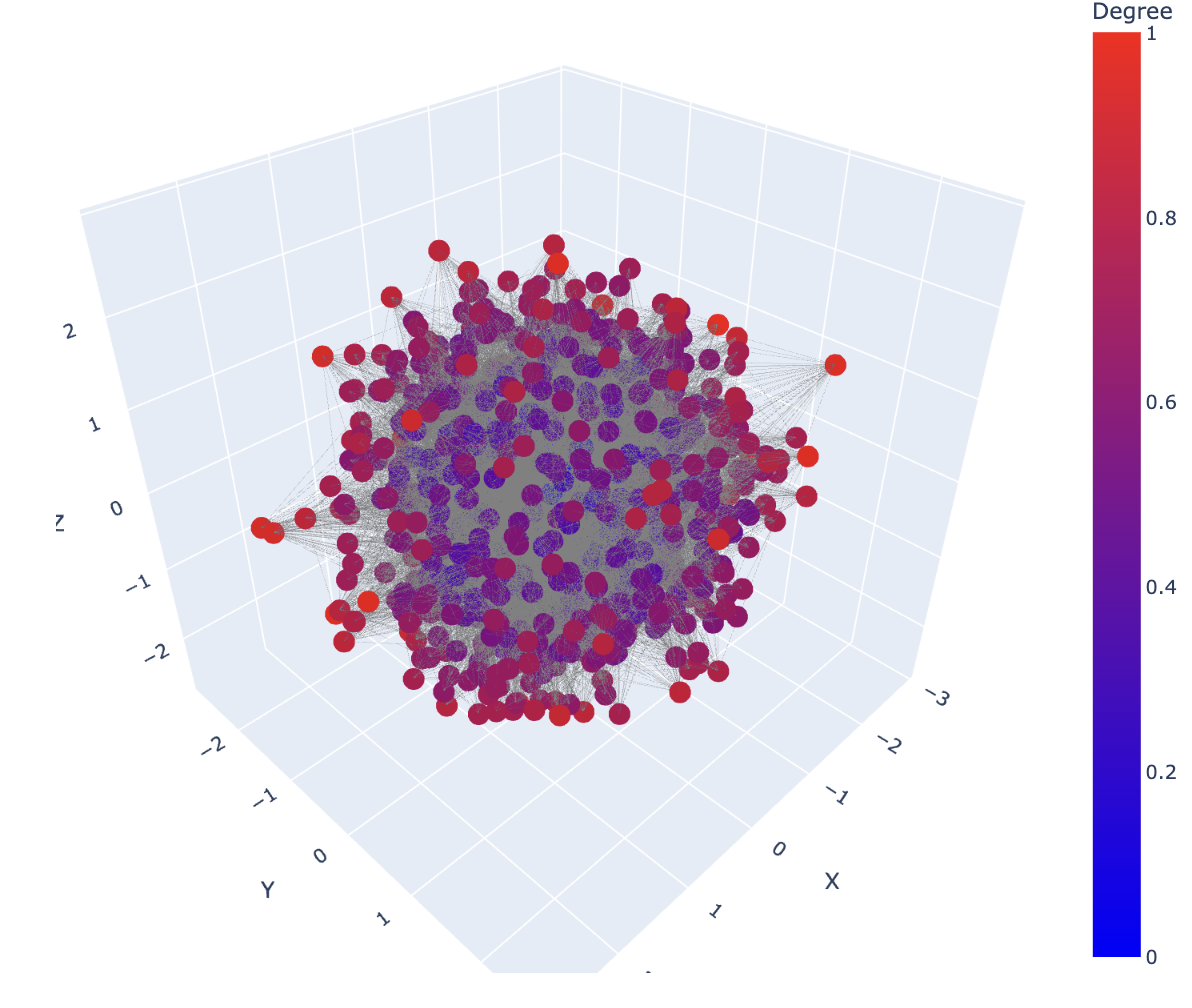}%
        \label{fig:layout_finished}
    }
    \caption{The initial embedding and the final layout. Vertex degrees (normalized to $[0,1]$) are shown in color: blue = lower degree, red = higher degree. Here an Erd\"os--Renyi graph is used with $n=1000$ vertices, and edge probability $p=0.025$.}
    \label{fig:er_layout}
\end{figure}

\begin{figure}[htbp]
    \centering
    \subfloat[Initial Laplacian Embedding]{%
        \includegraphics[width=0.45\linewidth]{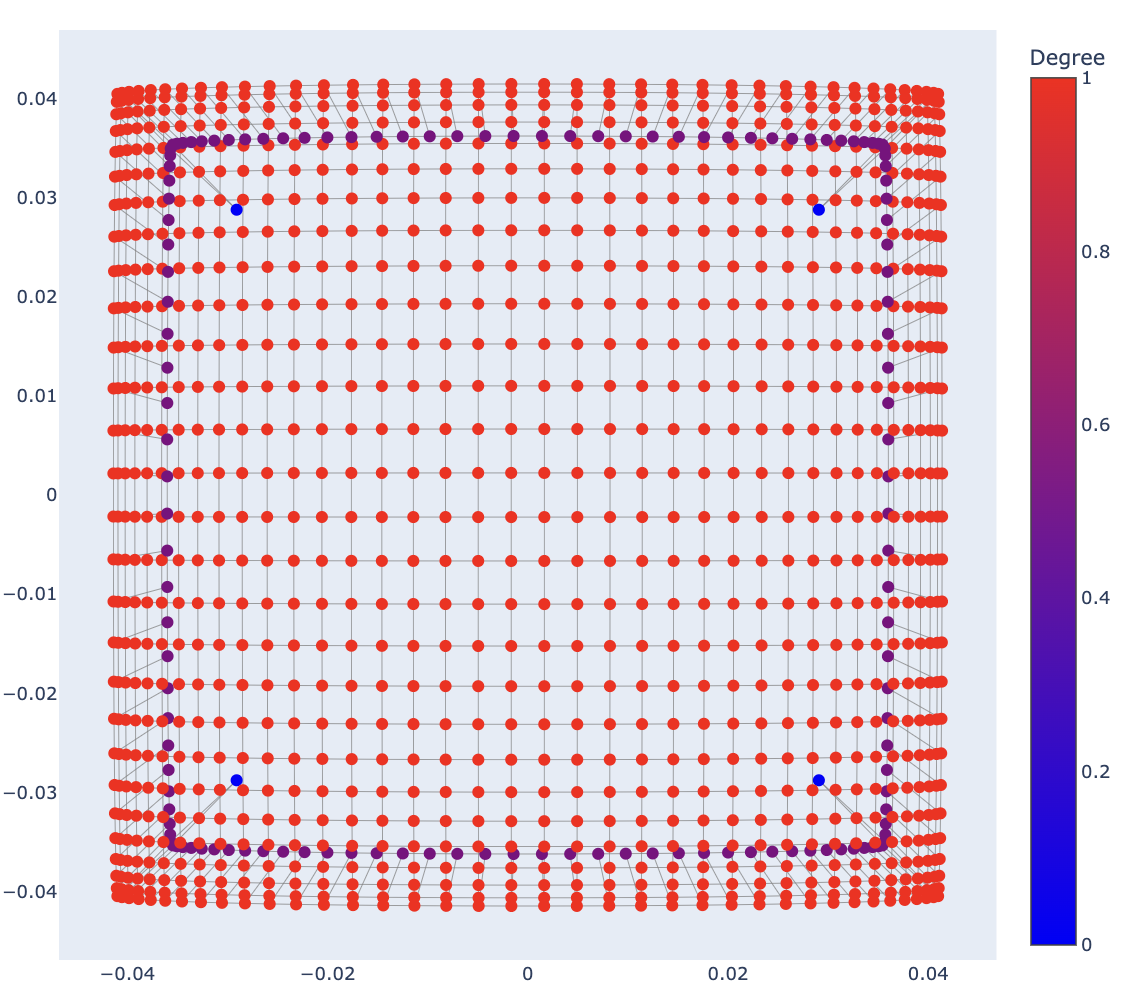}%
        \label{fig:grid_initial_embed}
    }
    \hfill
    \subfloat[Finalized Force Layout]{%
        \includegraphics[width=0.45\linewidth]{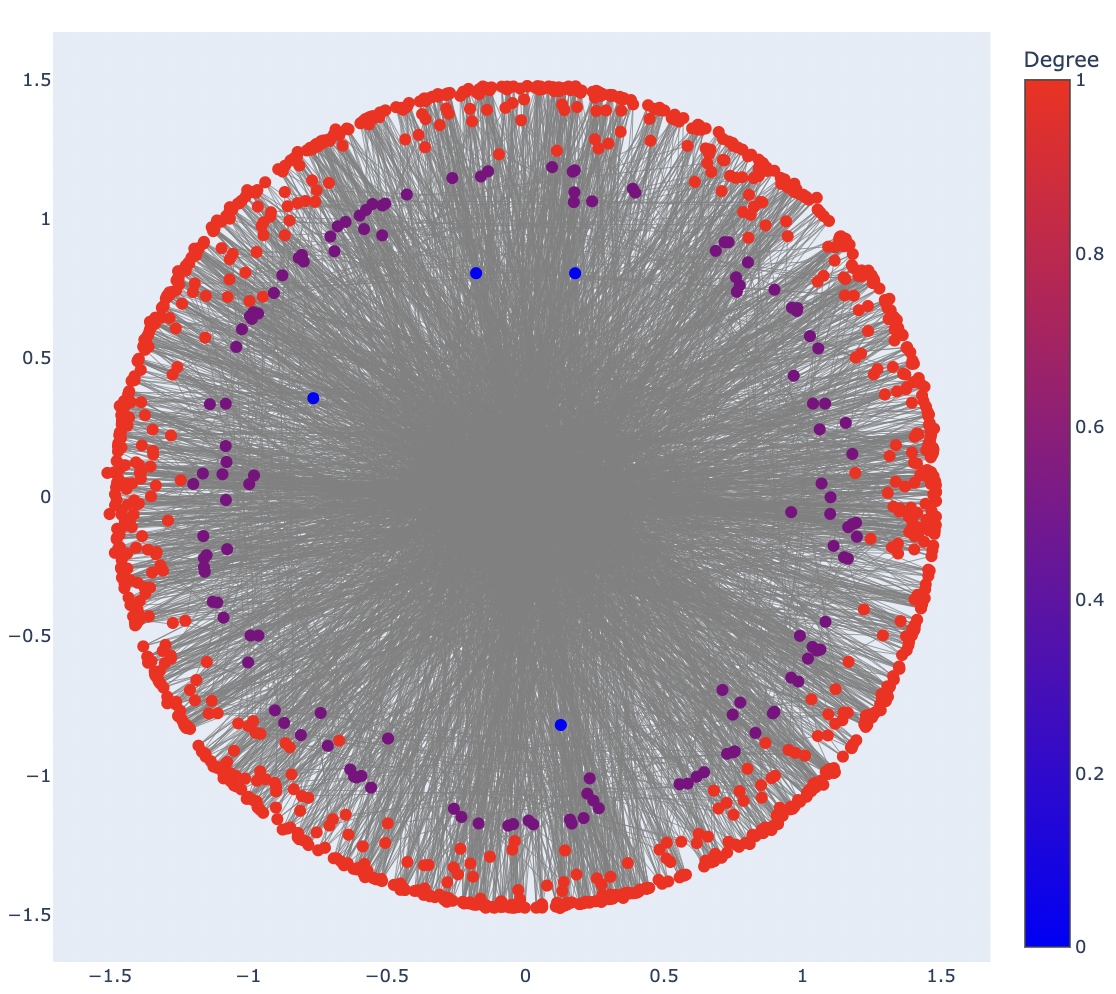}%
        \label{fig:grid_layout_finished}
    }
    \caption{The initial embedding and the final layout. Vertex degrees (normalized to $[0,1]$) are shown in color: blue = lower degree, red = higher degree. Here the grid graph with $30 \times 40$ squares is used. }
    \label{fig:grid_layout}
\end{figure}

\begin{figure}[htbp]
    \centering
    \subfloat[Initial Laplacian Embedding]{%
        \includegraphics[width=0.45\linewidth]{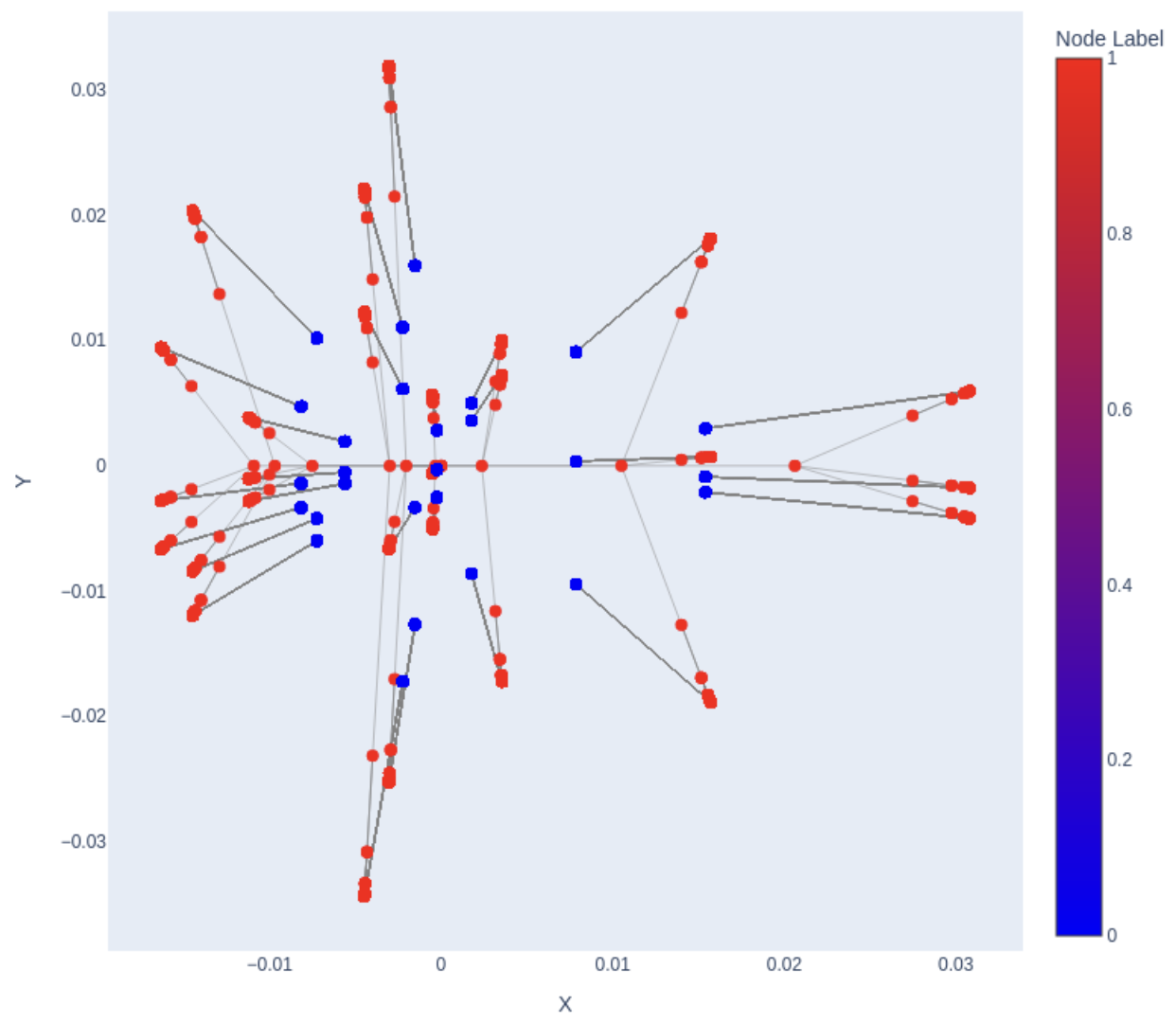}%
        \label{fig:tree_initial_embed}
    }
    \hfill
    \subfloat[Finalized Force Layout]{%
        \includegraphics[width=0.45\linewidth]{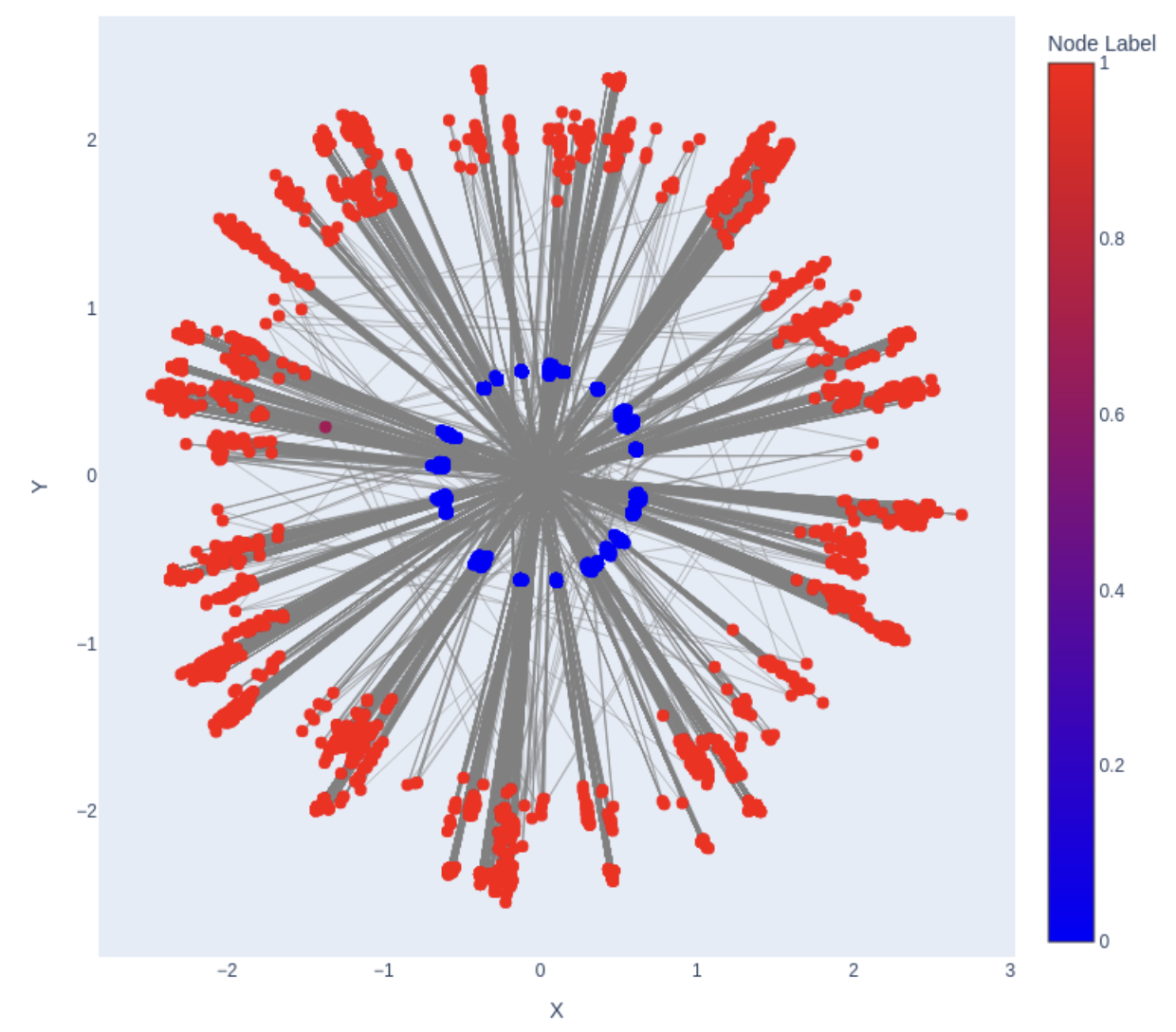}%
        \label{fig:tree_layout_finished}
    }
    \caption{The initial embedding and the final layout. Vertex degrees (normalized to $[0,1]$) are shown in color: blue = lower degree, red = higher degree. Here the balanced ternary tree with $8$ levels is used. A few vertices and edges appear to coincide in the initial embedding.}
    \label{fig:tree_layout}
\end{figure}

We also present some visual ``before--and--after'' images of the initial Laplacian embeddings of graphs in dimensions $2$ and $3$ (``before'') and their finalized force layout embeddings (``after''). Each embedding has the vertices color--labeled according to the vertex degree normalize to the interval from $0$ (minimum degree) to $1$ (maximum degree). One can observe that how high--centrality vertices are moved towards the periphery in the case of a random Erd\"os--Renyi (Figure~\ref{fig:er_layout}), a grid graph (Figure~\ref{fig:grid_layout}), and a balanced ternary tree (Figure~\ref{fig:tree_layout}). 

\subsection{Real World Datasets}

We use several dataset from SNAP \cite{snapnets} in order to test the algorithm's effectiveness on large scale real world data: 

\begin{itemize}
    \item ``General Relativity and Quantum Cosmology collaboration network'' \cite{snap-grqc, snap-grqc-paper} in Table~\ref{tab:snap_grqc};
    \item ``Social Circles: Facebook'' \cite{snap-facebook, snap-facebook-paper} in Table~\ref{tab:snap_facebook_combined};
    \item ``Wikipedia vote network'' \cite{snap-wiki-vote, snap-wiki-vote-paper}, with 74\% of the vertices (and 55\% of all edges) being subsampled to reduce the complexity of computation for combinatorial centrality measures, in Table~\ref{tab:snap_wiki_vote}. Degree, eigenvector and PageRank centralities can be computed for the entire dataset, and the resulting correlations are shown in Table~\ref{tab:snap_wiki_vote_complete}. 
\end{itemize}

For disconnected graphs we use the largest connected component for benchmarking. For very large graphs, since computing combinatorial centralities may not be feasible, we use subsampling. We also compare three embedding dimensions for our benchmarks, $d = 3$, $4$, and $6$, and record the best one. More benchmarks are available on GitHub \cite{graphem-github}.

\begin{table}[h!]
\centering
\begin{tabular}{lccc}
\hline
\textbf{Centrality Measure} & \boldmath$\rho$ & \textbf{95\% CI} & \boldmath$p$ \\
\hline
Degree       & 0.834 & [0.817, 0.850] & $< 10^{-5}$ \\
Betweenness  & 0.770 & [0.757, 0.785] & $< 10^{-5}$ \\
Eigenvector  & 0.370 & [0.344, 0.395] & $< 10^{-5}$ \\
PageRank     & 0.839 & [0.826, 0.853] & $< 10^{-5}$ \\
Closeness    & 0.450 & [0.422, 0.477] & $< 10^{-5}$ \\
Node Load    & 0.770 & [0.755, 0.784] & $< 10^{-5}$ \\
\hline
\end{tabular}
\caption{Spearman correlations of centrality measures with the radial distance in a graph embedding for the SNAP ``General Relativity and Quantum Cosmology collaboration network'' dataset \cite{snap-grqc}. Embedding dimension $6$.}
\label{tab:snap_grqc}
\end{table}

\begin{table}[h!]
\centering
\begin{tabular}{lccc}
\hline
\textbf{Centrality Measure} & \boldmath$\rho$ & \textbf{95\% CI} & \boldmath$p$ \\
\hline
Degree       & 0.864 & [0.851, 0.877] & $< 10^{-5}$ \\
Betweenness  & 0.721 & [0.704, 0.740] & $< 10^{-5}$ \\
Eigenvector  & 0.537 & [0.513, 0.560] & $< 10^{-5}$ \\
PageRank     & 0.746 & [0.730, 0.763] & $< 10^{-5}$ \\
Closeness    & 0.592 & [0.571, 0.610] & $< 10^{-5}$ \\
Node Load    & 0.718 & [0.698, 0.736] & $< 10^{-5}$ \\
\hline
\end{tabular}
\caption{Spearman correlations of centrality measures with the radial distance in a graph embedding for the SNAP ``Social circles: Facebook'' dataset \cite{snap-facebook}. Embedding dimension $4$.}
\label{tab:snap_facebook_combined}
\end{table}

\begin{table}[h!]
\centering
\begin{tabular}{lccc}
\hline
\textbf{Centrality Measure} & \boldmath$\rho$ & \textbf{95\% CI} & \boldmath$p$ \\
\hline
Degree       & 0.955 & [0.950, 0.959] & $< 10^{-5}$ \\
Betweenness  & 0.934 & [0.928, 0.939] & $< 10^{-5}$ \\
Eigenvector  & 0.852 & [0.840, 0.863] & $< 10^{-5}$ \\
PageRank     & 0.952 & [0.947, 0.956] & $< 10^{-5}$ \\
Closeness    & 0.839 & [0.827, 0.850] & $< 10^{-5}$ \\
Node Load    & 0.933 & [0.928, 0.938] & $< 10^{-5}$ \\
\hline
\end{tabular}
\caption{Spearman correlations of centrality measures with the radial distance in a graph embedding for the SNAP ``Wikipedia vote network'' dataset \cite{snap-wiki-vote}. Embedding dimension $3$. We subsampled $5250$ vertices to reduce combinatorial complexity.}
\label{tab:snap_wiki_vote}
\end{table}

\begin{table}[h!]
\centering
\begin{tabular}{lccc}
\hline
\textbf{Centrality Measure} & \boldmath$\rho$ & \textbf{95\% CI} & \boldmath$p$ \\
\hline
Degree       & 0.964 & [0.961, 0.967] & $< 10^{-5}$ \\
Eigenvector  & 0.871 & [0.863, 0.879] & $< 10^{-5}$ \\
PageRank     & 0.958 & [0.954, 0.962] & $< 10^{-5}$ \\
\hline
\end{tabular}
\caption{Spearman correlations of centrality measures with the radial distance in a graph embedding for the SNAP ``Wikipedia vote network'' dataset \cite{snap-wiki-vote}. Complete dataset with $7 115$ vertices and $103 689$ edges. Embedding dimension~$3$.}
\label{tab:snap_wiki_vote_complete}
\end{table}

\section{Node Influence Optimization}

Influence optimization is a central task in such domains as viral marketing, mathematical epidemiology, and social network analysis. While traditional greedy algorithms iteratively select nodes based on their highest marginal gains in influence, this approach is often computationally expensive. 

\subsection{Synthetic Dataset}

In our experiments, random Erd\"os--Renyi graphs were used with $n=128$ nodes and edge probability $p=0.05$. We used the Independent Cascades (IC) method with adjacent node activation probability $p_{ic} = 0.1$ and $k=10$ seed vertices. The IC algorithm was realized by using NDlib library \cite{ndlib}. The benchmark was repeated $50$ time to collect a statistical sample. The outcomes of using the graph embedding as opposed to the classical greedy seed selection are given in Table~\ref{tab:graphem_vs_greedy_synthetic} for comparison. 

\begin{table}[h!]
\centering
\begin{tabular}{lccc}
\hline
\textbf{Method} & \textbf{Influence} & \textbf{Iterations} & \textbf{Time (s)} \\
\hline
Embedding & $24.6 \pm 6.9$ & $200$ & $0.26 \pm 0.48$ \\
Greedy    & $23.7 \pm 5.1$ & $247 200$ & $15.97 \pm 0.08$ \\
\hline
\end{tabular}
\caption{Influence spread, number of NDlib simulation iterations, and runtime for embedding-based vs. greedy method. Synthetic dataset: a random Erd\"os-Renyi graph on $128$ nodes with edge probability $p=0.05$.}
\label{tab:graphem_vs_greedy_synthetic}
\end{table}

\subsection{Real World Dataset}

As a real world dataset, we use the ``General Relativity and Quantum Cosmology collaboration network'' dataset \cite{snap-grqc}. Its larges connected component has $4158$ nodes and $13428$ edges, which is within reasonable bounds to run greedy search for maximum influence subsets (while the graph embedding would work efficiently for much larger datasets). The benchmarking results are given in Table~\ref{tab:graphem_vs_greedy_real-world}. 

\begin{table}[h!]
\centering
\begin{tabular}{lccc}
\hline
\textbf{Method} & \textbf{Influence} & \textbf{Iterations} & \textbf{Time (s)} \\
\hline
Embedding & $23.9 \pm 6.0$ & $200$ & $0.19 \pm 0.01$ \\
Greedy    & $22.9 \pm 5.7$ & $247{,}200$ & $15.95 \pm 0.07$ \\
\hline
\end{tabular}
\caption{Influence spread, number of NDlib simulation iterations, and runtime for embedding-based vs. greedy method. Real world dataset: ``General Relativity and Quantum Cosmology collaboration network'' \cite{snap-grqc}.}
\label{tab:graphem_vs_greedy_real-world}
\end{table}

\section{Other embeddings}

\begin{figure}[htbp]
    \centering
    \subfloat[Laplacian Embedding]{%
        \includegraphics[width=0.45\linewidth]{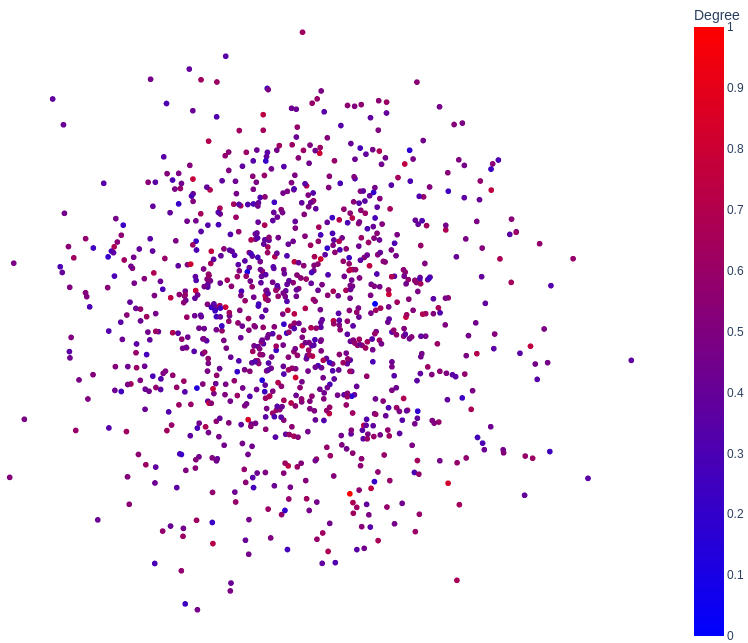}%
        \label{fig:laplacian_embedding_corr}
    }
    \hfill
    \subfloat[UMAP Embedding]{%
        \includegraphics[width=0.45\linewidth]{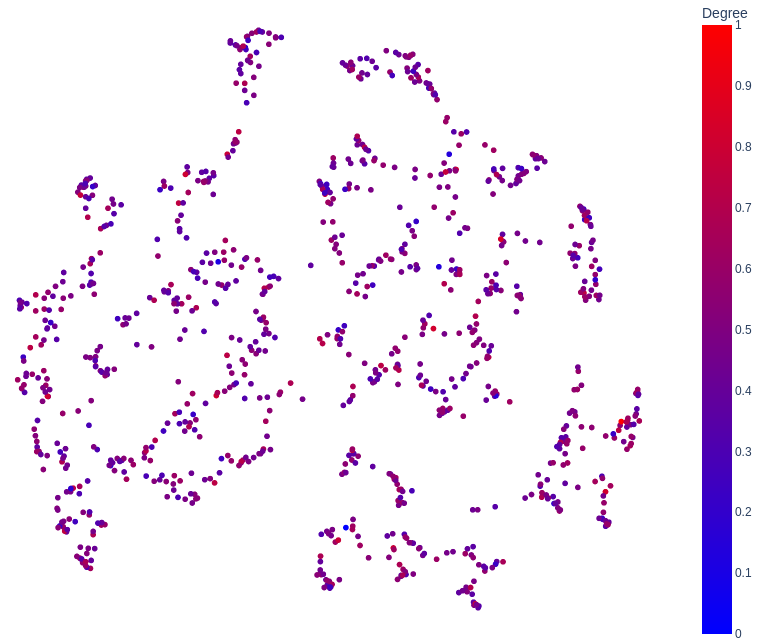}%
        \label{fig:umap_embedding_corr}
    }
    \hfill
    \subfloat[TriMAP Embedding]{%
        \includegraphics[width=0.45\linewidth]{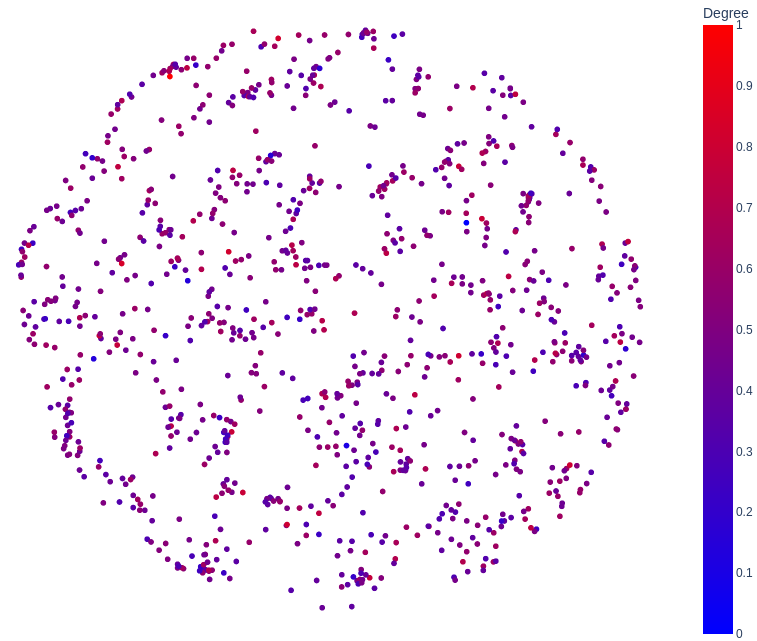}%
        \label{fig:trimap_embedding_corr}
    }
    \hfill
    \subfloat[PaCMAP Embedding]{%
        \includegraphics[width=0.45\linewidth]{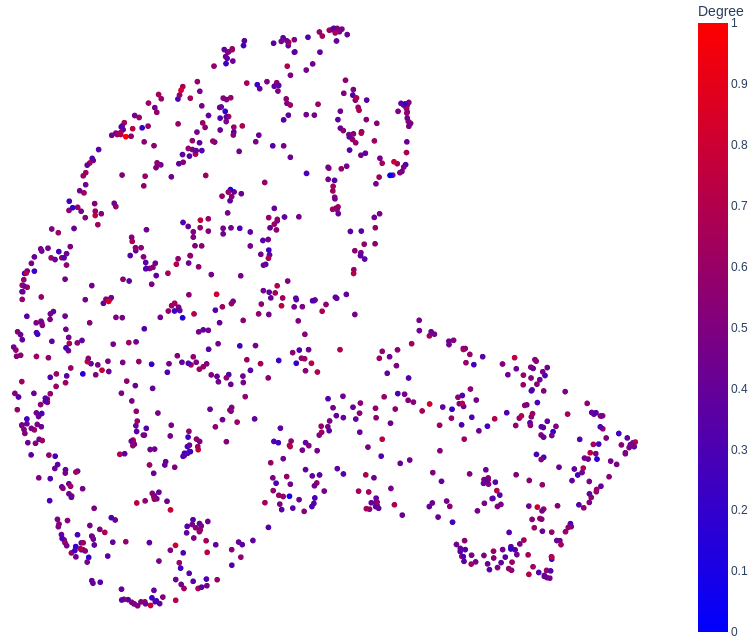}%
        \label{fig:pacmap_embedding_corr}
    }
    \caption{Graph layout produced by the Laplacian eigenmap, UMAP, TriMAP, and PaCMAP. The input is a random Erd\"os--Renyi graph with $1000$ vertices and edge probability $0.05$. Vertex degrees (normalized to $[0,1]$) are shown in color: blue = lower degree, red = higher degree.}
    \label{fig:other_graph_embeddings}
\end{figure}

In this section we check if other graph embeddings produce radial distance correlated with centrality measures. We produced embeddings in dimension $d=2$ by the following widely used methods: Laplacian eigenmap, UMAP \cite{umap, umap-git}, TriMAP \cite{amid2019trimap}, PaCMAP \cite{wang2021understanding, pacmap-git}. As inputs, we used random Erd\"os--Renyi graphs on $1000$ vertices with vertex probability $0.05$. There was no statistically significant Spearman correlation observed in any of the considered cases, see Tables~\ref{tab:laplacian_corr_centrality}, \ref{tab:umap_corr_centrality}, \ref{tab:trimap_corr_centrality}, and \ref{tab:pacmap_corr_centrality}. A visual inspection confirms the findings, as evidenced in Figure~\ref{fig:other_graph_embeddings}. 

\begin{table}[h!]
\centering
\begin{tabular}{lccc}
\hline
\textbf{Centrality Measure} & \boldmath$\rho$ & \textbf{95\% CI} & \boldmath$p$ \\
\hline
Degree       & $-0.018$ & [$-0.079$, $0.047$] & 0.563 \\
Betweenness  & $-0.026$ & [$-0.078$, $0.033$] & 0.408 \\
Eigenvector  & $-0.025$ & [$-0.084$, $0.037$] & 0.428 \\
PageRank     & $-0.016$ & [$-0.078$, $0.047$] & 0.605 \\
Closeness    & $-0.050$ & [$-0.109$, $0.023$] & 0.111 \\
Node Load    & $-0.027$ & [$-0.090$, $0.035$] & 0.400 \\
\hline
\end{tabular}
\caption{Laplacian eigenmaps and centrality measures {\it do not} exhibit statistically significant Spearman correlation.}
\label{tab:laplacian_corr_centrality}
\end{table}

\begin{table}[h!]
\centering
\begin{tabular}{lccc}
\hline
\textbf{Centrality Measure} & \boldmath$\rho$ & \textbf{95\% CI} & \boldmath$p$ \\
\hline
Degree       & $-0.009$ & [$-0.069$, $0.046$] & 0.765 \\
Betweenness  & $-0.020$ & [$-0.080$, $0.042$] & 0.532 \\
Eigenvector  & $-0.013$ & [$-0.080$, $0.047$] & 0.679 \\
PageRank     & $-0.008$ & [$-0.077$, $0.054$] & 0.790 \\
Closeness    & $-0.038$ & [$-0.101$, $0.024$] & 0.235 \\
Node Load    & $-0.020$ & [$-0.083$, $0.041$] & 0.519 \\
\hline
\end{tabular}
\caption{UMAP embedding and centrality measures {\it do not} exhibit statistically significant Spearman correlation.}
\label{tab:umap_corr_centrality}
\end{table}

\begin{table}[h!]
\centering
\begin{tabular}{lccc}
\hline
\textbf{Centrality Measure} & \boldmath$\rho$ & \textbf{95\% CI} & \boldmath$p$ \\
\hline
Degree       & $-0.008$ & [$-0.072$, $0.057$] & 0.811 \\
Betweenness  & $-0.017$ & [$-0.075$, $0.049$] & 0.587 \\
Eigenvector  & $-0.012$ & [$-0.077$, $0.051$] & 0.694 \\
PageRank     & $-0.006$ & [$-0.072$, $0.054$] & 0.844 \\
Closeness    & $-0.038$ & [$-0.098$, $0.022$] & 0.231 \\
Node Load    & $-0.018$ & [$-0.079$, $0.045$] & 0.574 \\
\hline
\end{tabular}
\caption{TriMAP embedding and centrality measures {\it do not} exhibit statistically significant Spearman correlation.}
\label{tab:trimap_corr_centrality}
\end{table}

\begin{table}[h!]
\centering
\begin{tabular}{lccc}
\hline
\textbf{Centrality Measure} & \boldmath$\rho$ & \textbf{95\% CI} & \boldmath$p$ \\
\hline
Degree       & $-0.023$ & [$-0.081$, $0.038$] & 0.462 \\
Betweenness  & $-0.029$ & [$-0.096$, $0.039$] & 0.363 \\
Eigenvector  & $-0.031$ & [$-0.089$, $0.037$] & 0.333 \\
PageRank     & $-0.021$ & [$-0.079$, $0.041$] & 0.507 \\
Closeness    & $-0.051$ & [$-0.101$, $0.009$] & 0.104 \\
Node Load    & $-0.029$ & [$-0.089$, $0.035$] & 0.355 \\
\hline
\end{tabular}
\caption{PaCMAP embedding and centrality measures {\it do not} exhibit statistically significant Spearman correlation.}
\label{tab:pacmap_corr_centrality}
\end{table}

\section{Discussion}

We produce a new graph embedding method that provides a proxy for node centrality in a wide variety of graph classes, as is evidenced by our numerical experiments on both synthetic and real world datasets. The method's convergence and properties have a mathematical backing, although some assumptions should be made. The heuristic arguments  

Other graph embedding methods, of similar kin, do not exhibit strong statistical correlation between the radial distance to the embedded nodes and their importance, even in relatively common and simple classes of graphs such as random Erd\"os--Renyi graphs. 

This embedding may be potentially generalized and enhanced to the case of directed graphs, although we leave it as a possible direction for future work. 

\section*{Code and data availability}

The code used in the present paper is available on GitHub \cite{graphem-github}. All necessary data is either generated by this code (synthetic datasets), or downloaded from SNAP \cite{snapnets} (real world datasets). 

\section*{Acknowledgments} This work is supported by the Google~Cloud Research Award number GCP19980904.

\printbibliography

\end{document}